\documentclass[aps,prl,twocolumn,superscriptaddress,showpacs]{revtex4-1}
\usepackage{graphics}
\usepackage{graphicx}
\usepackage{amsmath}
\usepackage{amsfonts}
\usepackage{bbm,bm}

\ifx\pdftexversion\undefined
\usepackage[dvips]{hyperref}
\else
\usepackage[colorlinks, allcolors=blue]{hyperref}
\fi
\hypersetup{
  colorlinks = true, linkcolor = blue
}

\begin{document}

\title{Certifying the adiabatic preparation of ultracold lattice bosons in the vicinity of the Mott transition}

\author{C\'ecile Carcy}
\affiliation{Universit\'e Paris-Saclay, Institut d'Optique Graduate School, CNRS, Laboratoire Charles Fabry, 91127, Palaiseau, France}
\author{Ga\'etan Herc\'e}
\affiliation{Universit\'e Paris-Saclay, Institut d'Optique Graduate School, CNRS, Laboratoire Charles Fabry, 91127, Palaiseau, France}
\author{Antoine Tenart}
\affiliation{Universit\'e Paris-Saclay, Institut d'Optique Graduate School, CNRS, Laboratoire Charles Fabry, 91127, Palaiseau, France}
\author{Tommaso Roscilde}
\affiliation{Universit\'e de Lyon, Ens de Lyon, Univ. Claude Bernard and CNRS, Laboratoire de Physique, F-69342 Lyon, France}
\author{David Cl\'ement}
\affiliation{Universit\'e Paris-Saclay, Institut d'Optique Graduate School, CNRS, Laboratoire Charles Fabry, 91127, Palaiseau, France}
\date{\today}

\begin{abstract}
We present a joint experimental and theoretical analysis to assess the adiabatic experimental preparation of ultracold bosons in optical lattices aimed at simulating the three-dimensional Bose-Hubbard model. Thermometry of lattice gases is realized from the superfluid to the Mott regime by combining the measurement of three-dimensional momentum-space densities with ab-initio quantum Monte Carlo (QMC) calculations of the same quantity. The measured temperatures are in agreement with isentropic lines reconstructed via QMC for the experimental parameters of interest, with a conserved entropy per particle of $S/N=0.8(1) k_{B}$. In addition, the Fisher information associated with this thermometry method shows that the latter is most accurate in the critical regime close to the Mott transition, as confirmed in the experiment.  These results prove that equilibrium states of the Bose-Hubbard model -- including those in the quantum-critical regime above the Mott transition -- can be adiabatically prepared in cold-atom apparatus. 
\end{abstract}

\maketitle 

The simulation of strongly interacting quantum systems in experiments represents a most promising research effort, relying on the exquisite level of control acquired on different platforms -- from ultracold atoms \cite{Blochetal2008} to semiconducting  \cite{Hensgensetal2017} or superconducting \cite{Roushanetal2017} circuits. When the goal is the realization of an equilibrium state of a quantum many-body system, a paradigm common to all these platforms is that of adiabatic preparation \cite{AlbashL2018,Haukeetal2020} in the absence of an external heat bath: starting from a fiducial quantum state of an initial Hamiltonian, a continuous variation of the Hamiltonian parameters aims at transforming the state into the equilibrium state of a target Hamiltonian, at constant entropy. The successful implementation of the above paradigm is yet far from obvious, and depends on whether the quantum-state preparation is performed at (nearly) zero entropy or at finite entropy. 

In platforms manipulating small ensembles ($N \lesssim 10^2-10^3$) of discrete quantum variables (akin to quantum spins) -- such as trapped ions \cite{Monroeetal2014}, Rydberg atoms \cite{BrowaeysL2020} and quantum circuits \cite{Kingetal2018} -- the initial state can be prepared as the (nearly) pure ground state at (close to) zero entropy. The conditions for its adiabatic transformation upon varying the Hamiltonian are mostly dictated by the size of the gap to the excited states \cite{AlbashL2018}. The main obstacle to this \emph{pure-state adiabaticity} is therefore offered by the vanishing of the excitation gap upon increasing the system size, \emph{e.g.} at a quantum phase transition. 

The situation is different in the case of quantum simulators where finite-entropy states are manipulated, such as with ultracold bosons \cite{Blochetal2008}. There, typical experiments start from a non-zero entropy state (a Bose-Einstein condensate at finite temperature) of a large number of degrees of freedom ($N\sim 10^3-10^5$), followed by a transformation of the system's Hamiltonian (such as the loading of atoms in an optical lattice). Extending naively the criteria of \emph{pure-state adiabaticity} to a mixed state would suggest prohibitive conditions, as the energy gaps in the middle of the spectrum are exponentially small in the system size. Instead \emph{mixed-state adiabaticity} does not require to follow adiabatically each pure state of the mixture, but rather to produce a state compatible with an equilibrium state of the instantaneous Hamiltonian at the same entropy. What are the conditions to guarantee such \emph{mixed-state adiabaticity}? And what is the effect of quantum phase transitions (occurring in the ground state) on a finite-entropy transformation?

These are in fact formidable questions, that are being theoretically addressed only recently \cite{ilinetal2020}, and which are potentially very hard to answer to with unbiased calculations. In experiments (such as those with lattice Bose gases), keeping the entropy at a low value upon Hamiltonian transformations has always been a central preoccupation \cite{Gerickeetal2007}. But quantitative answers to the above questions are missing, mostly because a direct measure of the entropy in the experiment is hardly accessible. On the other hand, a viable route to probe the adiabatic preparation of complex many-body quantum states at finite entropy results from the combination of experiments with ab-initio calculations \cite{Trotzkyetal2010, Caylaetal2018}. Indeed experimental (quasi-) adiabatic processes can be certified whenever the expected equilibrium state produced by the evolution can be efficiently simulated classically (using \emph{e.g.} quantum Monte Carlo). This program of certifying finite-entropy adiabatic processes, including the crossing of a quantum phase transition, is precisely the object of this work. 
 
\begin{figure}[t!]
\includegraphics[width=\columnwidth]{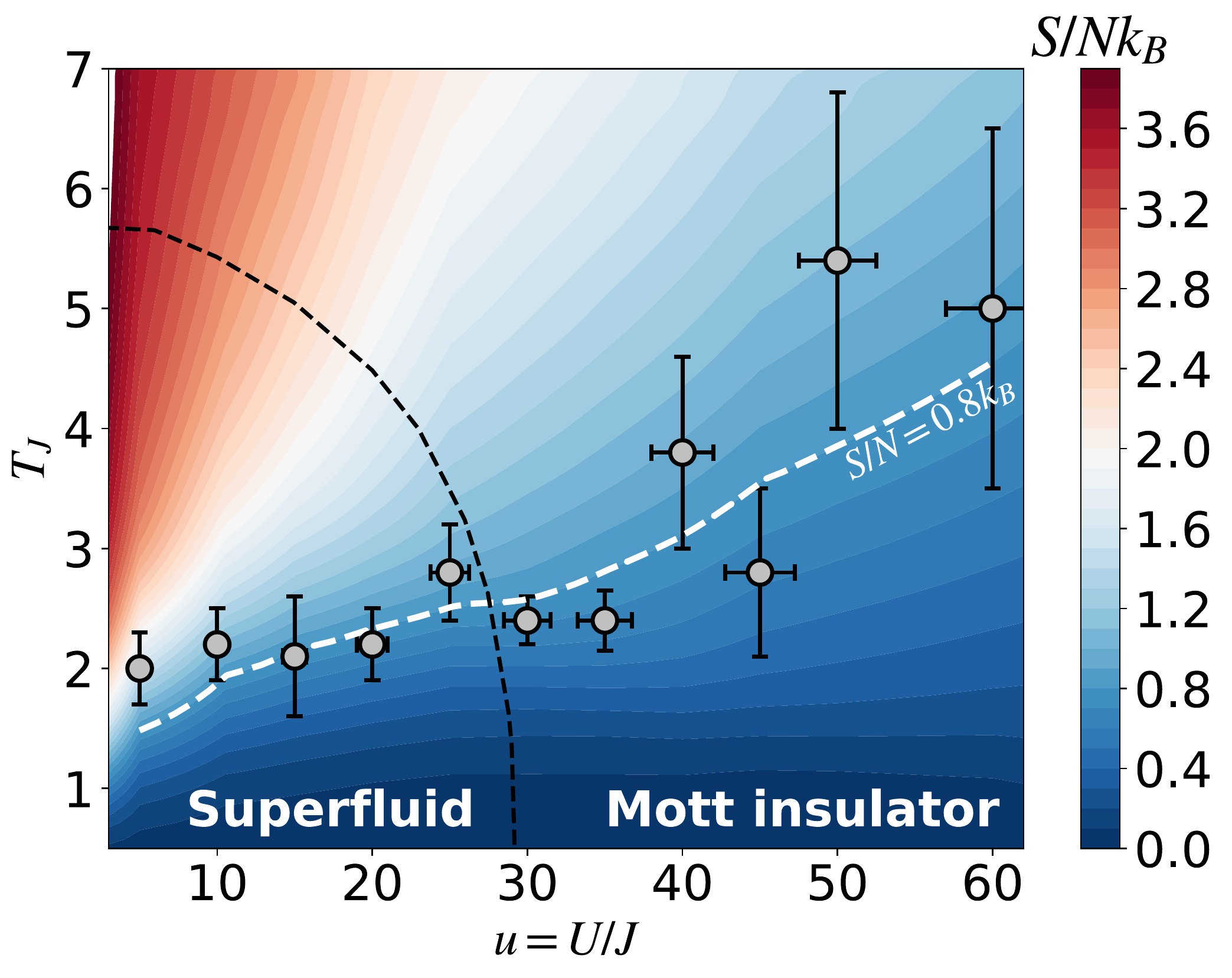}
\caption{Experimental reduced temperatures $T_{J}=k_{B} T /J$ (open dots) obtained from the momentum-space-density thermometry (see text) plotted as a function of the ratio $u=U/J$, and compared to isentropic lines. The underlying false-color plot shows the theoretical map of the entropy per particle $S/N k_{B}$ of the trapped 3D BH model, with the same parameters (particle number, interaction strength, trapping potential) as in the experiment. The white dashed curve (isentropic line at $S/N = 0.8 k_B$) offers the best agreement with the experimental data. The black dashed line, shown for reference, represents the line of critical temperatures for the uniform 3D BH model at unit filling (from Ref.~\cite{Capogrosso-Sansoneetal2007}).}
\label{f.thermometry}
\end{figure}
   
In this Letter, we focus our attention on the adiabatic preparation of low-energy equilibrium states in the three-dimensional (3D) Bose-Hubbard (BH) model
 \begin{equation}
 {\cal H} = - J \sum_{\langle ij \rangle} \left ( b_i^\dagger b_j + {\rm h.c.} \right) + \sum_i \left [ \frac{U}{2} n_i (n_i -1) + V_i n_i \right]
 \label{e.3dBH}
 \end{equation}
 where $b_i, b_i^\dagger$ and $n_i = b_i^\dagger b_i$ are bosonic operators, $\langle ij \rangle$ are nearest-neighbor pairs on the cubic lattice, $J$ is the hopping amplitude and $U$ the on-site repulsive interaction energy, and $V_i = V_x x_i ^2 + V_y y_i ^2 + V_z z_i ^2$ is an overall parabolic trapping potential.  When working at uniform integer filling $n$ this model is known to possess a ground-state quantum phase transition from a gapless superfluid (SF) phase to a gapped Mott insulator (MI) phase upon increasing the ratio $u=U/J$ - in the case $n=1$ the transition is estimated to occur for $u_c = 29.34$ \cite{Capogrosso-Sansoneetal2007}. We implement the physics of the 3D BH model using interacting bosons of metastable Helium-4 atoms ($^4$He$^*$) loaded in a 3D optical lattice \cite{Caylaetal2018,Carcyetal2019}. The depth of the optical lattice sets the value of $u$ and provides a tool to cross the critical value $u_c$ for the SF/MI transition \cite{Greineretal2002}. Previous experiments \cite{Trotzkyetal2010} have demonstrated that slow ramps of the optical lattice produce interacting superfluid states (up to $u \lesssim u_{c}$). Yet the adiabatic nature of the loading process, and in particular the possible effect of the quantum phase transition on it (for $u>u_{c}$), remains to be tested. To achieve this goal, we exploit two ingredients associated with the detection of $^4$He$^*$ atoms: (i) the measurement of the 3D momentum-space density $\rho({\bm k})$ using multi-channel-plate detectors \cite{Nogrette2015}, offering the finest level of diagnostics on the first-order phase coherence; (ii) the single-atom sensitivity that permits the study of ensembles with a moderate number of atoms ($N\approx 3000$), with the benefit that ab-initio quantum Monte Carlo (QMC) simulations are achievable down to low temperatures. The combination of high-resolution measurements with ab-initio simulations allows us to extensively certify the preparation of equilibrium states of the 3D BH model (see below). In addition, we quantify the entropy per particle $S/N$ in the experiment, and we find that it is conserved as $u$ is varied (with a value $S/N\sim0.8 k_{B}$) even when crossing the critical value $u_{c}$. This conclusion, illustrated in Fig.~\ref{f.thermometry}, is the main result of our work.  
 
\begin{figure}[ht!]
\includegraphics[width=\columnwidth]{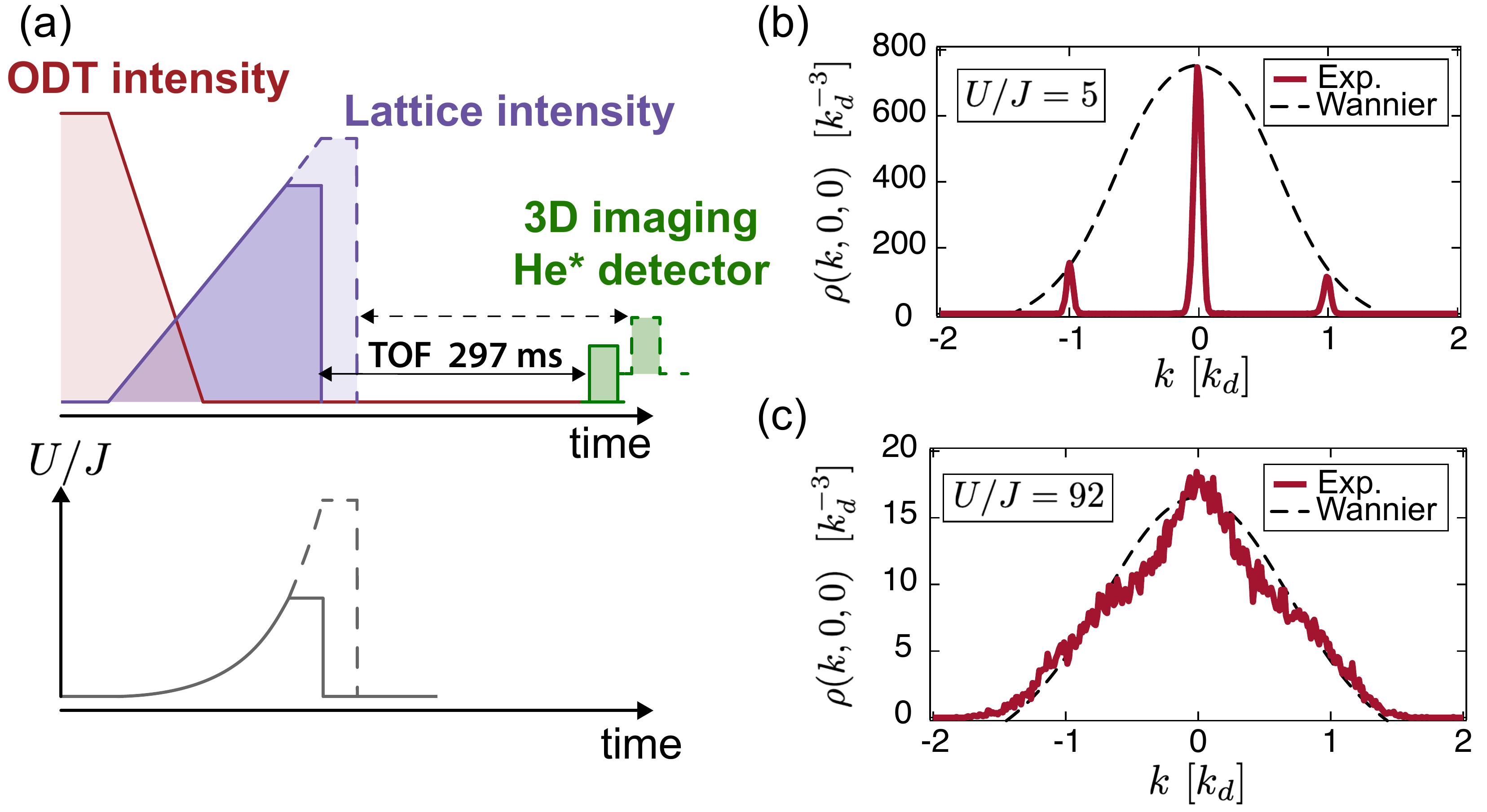}
\caption{(a) Time sequence for the loading of the $^4$He$^*$ BECs from the Optical Dipole Trap in the 3D optical lattice. The intensity of the 3D lattice is increased linearly with time at a rate of $0.3~E_{r}$.ms$^{-1}$. For two different final values of the lattice intensity, the two ramps coincide up to reaching the lowest of the two values. The linear increase of the lattice intensity corresponds to an approximately exponential increase of the ratio $U/J$ over time. (b)-(c) 1D cut $\rho({\bm k}=(k,0,0))$ along the $\vec{u}_{x}$ axis  through the 3D momentum-space densities measured at $u=5$ and $u=92$. }
\label{fig2}
\end{figure}

The experiments starts with the production of $^4$He$^*$ Bose-Einstein condensates (BECs) in a crossed Optical Dipole Trap (ODT) \cite{Boutonetal2015}. The BECs are then loaded into the lowest energy band of a 3D cubic optical lattice, characterized by a lattice spacing $d=775~$nm and an amplitude $V_0=s E_r$ where $E_r=h^2/8md^2$ is the lattice recoil energy \cite{Caylaetal2018,Tenartetal2020}. In the lattice potential, the overall harmonic trap is nearly isotropic with a frequency $140(10) \times \sqrt{s}$~Hz (see \cite{SuppMat} for detailed experimental parameters). The BEC atom number $N=3.0(5)\times 10^3$ used in this work ensures a lattice filling $n_{0}$ at the trap center equal or smaller than one atom per site, $n_{0}\lesssim1$. To load the atoms in the 3D lattice, $V_{0}$ is increased linearly at a rate of $0.3~E_{r}$/ms while the intensity of the ODT is decreased linearly to zero in $20~$ms  (see Fig.~\ref{fig2}(a)). The linear increase of $V_{0}$ corresponds to an almost exponential increase of $u$ (see Fig.~\ref{fig2}(a)). The shape and parameters of the ramps were optimized by reducing the heating and the atom losses observed after a protocol that transfers atoms in the lattice and back to the bare ODT. The ramps used to transfer the atoms from the lattice back to the ODT are the time reversal of those used to load atoms in the lattice. For the purposes of this work, it is important to note that probing the gas at a larger (final) value $u'>u$ is identical to starting from the equilibrium reached at $u$ and further increasing $V_{0}$ to reach $u'$. At the final lattice amplitude $V_{0}$, we hold the atoms for 5~ms before switching off the lattice potential abruptly (within 1~$\mu$s) and letting the gas expand. We then measure the 3D distribution of individual atoms with the He$^*$ detector after a time of flight (TOF) of $297~$ms \cite{Caylaetal2018}. 

We record 3D atom distributions at various amplitudes of the lattice across the SF-MI transition, spanning the ratio $u=U/J$ from $u=5$ to $u=92$. For each value $u$, the distribution results from averaging over about $M \sim 600$ runs of the experiment, and permits to extract the $k$-space density $\rho({\bm k})$, as well as the atom correlations \cite{Carcyetal2019, Caylaetal2020}. Two examples of profiles $\rho({\bm k}=(k,0,0))$ are shown in Fig.~\ref{fig2}(b)-(c). In contrast to previous works \cite{Xuetal2006, Trotzkyetal2010, Mackayetal2015}, we do not observe the presence of an incoherent background in the momentum-space densities. This probably derives from the difference in the detection methods: optical probes \cite{Xuetal2006, Trotzkyetal2010, Mackayetal2015} yield line-of-sight integrated 2D densities at moderate TOF durations \cite{Gerbieretal2008, Rayetal2013}, while the He$^*$ detector provides us with the 3D density in the far-field regime of expansion. 

\begin{figure}[ht!]
\includegraphics[width=\columnwidth]{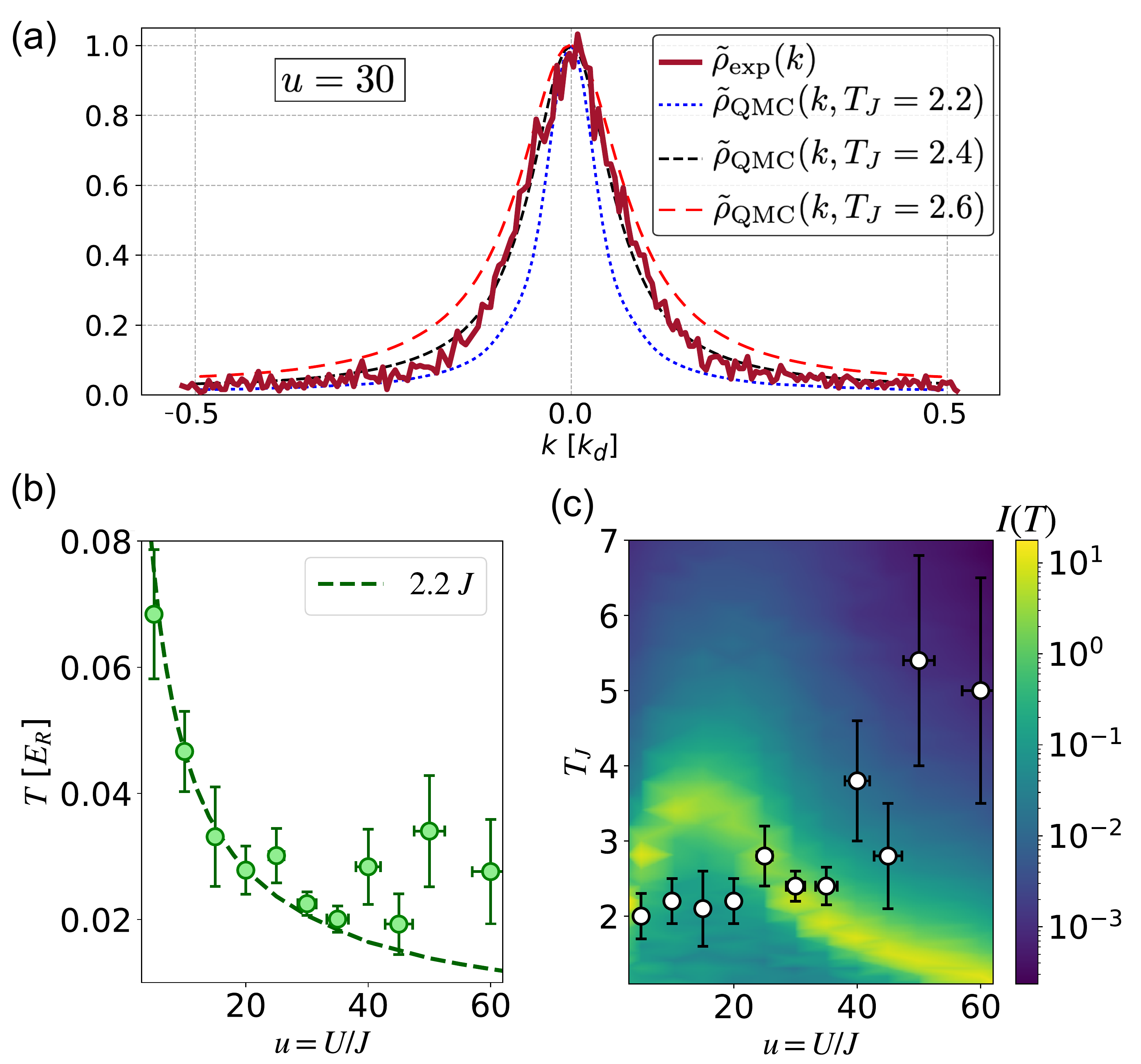}
\caption{(a) Plot of the momentum-space density $\tilde{\rho}_{\rm exp}(k)$ measured in the experiment (restricted to the first Brillouin zone) and of the theoretical one $\tilde{\rho}_{\rm QMC}(k;T)$ at various temperatures $T_{J}$. The comparison is shown for a ratio $u=30$ corresponding to the location of the quantum critical point in the ground-state; (b) Measured temperature $T$ in recoil units $E_{r}$ as a function of $u$. The solid line corresponds to $2.2~J$,  expressed in units of $E_{r}$; (c) Fisher information $I(T)$ for the temperature estimation from $\tilde\rho$ obtained from the QMC data; the experimental temperatures are reproduced for reference.}
\label{fig3}
\end{figure}

The temperature $T$ of the lattice gas can not be extracted directly from the measured momentum-space densities $\rho({\bm k})$ since an analytical prediction for the trapped 3D BH model of Eq.~\eqref{e.3dBH} does not exist. Instead, we use a thermometry method that relies on the fact that $\rho({\bm k})$ can be obtained ab-initio using quantum Monte Carlo (QMC) simulations. Since all the experimental parameters but the temperature are known, $T$ is the only adjustable parameter in the comparison with QMC simulations -- in particular we make use of Stochastic Series Expansion \cite{SyljuasenS2002} in the canonical ensemble \cite{Roscilde2008}, with a fixed particle number $N = 3000$. More specifically, $T$ is estimated as the temperature which minimises the distance between the measured normalized $k$-space density $\tilde{\rho}_{\rm exp}(k)=\rho(k,0,0)/\rho(0)$ with the theoretical one $\tilde{\rho}_{\rm QMC}(k;T)= \rho_{\rm QMC}(k,0,0;T)/\rho_{\rm QMC}(0;T)$ -- focusing on the momentum cut along ${\bm k} = (k,0,0)$.  Such a comparison relies on two assumptions that can only be verified a posteriori, by exhibiting a convincing agreement between the experimental and theoretical data: (i) the experiment realizes a thermal equilibrium state of the 3D BH model; (ii) the temperature of the equilibrium state is well defined in spite of the shot-to-shot fluctuations of the atom number $N$. The second assumption raises as well a question for the QMC calculations. In principle, the numerics should involve averaging at different atom numbers $N$, which is computationally rather demanding (in particular for the entropy calculations, see below). A detailed analysis of the effect of $N$ fluctuations (see \cite{SuppMat}) shows that such an average is not needed in practice. For the temperature and interaction regimes explored in the experiment, the quantity we use for the thermometry, namely $\tilde{\rho}_{\rm QMC}(k;T)$, shows a dependence on $N$ that spans a smaller range in densities than that associated with the experimental uncertainty. In other words, under the assumptions listed above, the experiment should reproduce (within its uncertainty range) results which are consistent with the equilibrium behavior for the 3D BH model in the canonical ensemble. For all the lattice depths, the theoretical density $\tilde{\rho}_{\rm QMC}(k)$ corresponding to the optimal temperature matches well the experimental density $\tilde{\rho}_{\rm exp}(k)$. This is the case even in the critical regime of the Mott transition, as illustrated in Fig.~\ref{fig3}(a) for $u=30$, for which the minimum reduced chi-square corresponding to the optimal temperature is compatible with unity, as shown in the Supp. Mat. \cite{SuppMat}. This justifies a posteriori our working assumptions. 

The results of the thermometry analysis are summarised in Fig.~\ref{f.thermometry}, which encompasses all the relevant regimes of the 3D BH model. In the SF regime, the reduced temperature $T_{J}= k_B T /J$ is estimated with a small uncertainty ($\sim 10-20\%$). In contrast, when the ground-state is a MI, a significant degradation is observed. We attribute this effect to the opening of an energy gap $\Delta$ in the excitation spectrum for the excitations localized in the center of the trap, suppressing thermal effects up to temperatures $T \gtrsim \Delta$: the experimental data, due to their finite precision, become compatible with theoretical densities that span a significantly larger interval in $T$. 
Note that the small increase of $T_{J}$ in the SF regime (see Fig.~\ref{f.thermometry}) should \emph{not} be associated with some heating mechanism. Indeed, when $T$ is expressed in absolute units, adiabatic cooling is observed (see Fig.~\ref{fig3}(b)). This descends from the fact that the isentropic gas is contained in a Bloch band whose width is proportional to $J$ and decreases with $s$.

To assess the precision of the thermometry, we compute numerically the Fisher information $I(T)$ (Fig.~\ref{fig3}(c)) that captures the sensitivity of $\tilde{\rho}(k)$ to variations in $T$, and which bounds the minimal uncertainty on the temperature obtained via the $k$-space thermometry \cite{SuppMat}, $(\delta T_J)_{\rm min} = [I(T) M]^{-1/2}$. We find that $I(T)$ takes its smallest values in the Mott insulator phase, consistent with the observed loss of accuracy in the experiment. In addition, for the parameters used in this work, $I(T)$ is maximal, and the temperature therefore best estimated, for $u \sim u_{c}$. The dramatic increase of $I(T)$  close to a phase transition reflects the critical increase of the Fisher information for the whole quantum state \cite{mehboudi2019}. The $k$-space thermometry, being optimal in the vicinity of the Mott transition, is therefore ideally suited to study the adiabatic character of state preparation above the quantum critical point $u\sim u_{c}$. Importantly, the error bars on the estimated temperature in the experiment exhibit a variation with $u$ compatible with that of $(\delta T_J)_{\rm min}$ set by the Fisher information. Near the optimal point $u \approx u_c$, the uncertainty on the estimated temperature is close to the theoretical limit $(\delta T_J)_{\rm min}$ \cite{SuppMat}. This demonstrates that our implementation of the $k$-space thermometry with $^4$He$^*$ can nearly saturate its maximum allowed precision.

We now turn to discussing the entropy of the lattice gases. Along with the $k$-space density, the QMC simulations yield the average energy per particle $e(T) = \langle {\cal H} \rangle/N$. A high-order polynomial fit to the energy allows one to extract the specific heat $c(T) = de(T)/dT$ and the entropy $S(T)/N = \int_0^T d\theta ~c(\theta)/\theta$ \cite{Capogrosso-Sansoneetal2007}. The QMC calculations can access the energy and specific heat of the trapped 3D BH model down to the lowest temperatures (required to reconstruct the entropy) thanks to the moderate particle number and system sizes explored in the experiment. In view of the above assumptions, the theoretically estimated entropy should reconstruct that of the thermal equilibrium state realized in the experiment.   
 
Fig.~\ref{f.thermometry} depicts the full entropy map of the trapped 3D BH model reconstructed with QMC over the temperature and interaction ranges relevant to the experiment. Besides the features of the entropy map -- which we shall comment below -- the most important observation that can be made is that all the experimental temperatures (except the one at $u=5$) are compatible with isentropic curves spanning the entropy range $S/N = 0.8(1)~k_B$. Within the uncertainty on the temperature, the experimental data are consistent with the picture in which the lattice ramp produces a sequence of thermal equilibrium states; and in which these states are connected by transformations conserving the entropy. This represents our strongest form of certification for the adiabatic preparation of equilibrium states of the 3D BH model in the experiment. In addition, the entropy of the lattice gas is compatible with the entropy $S_0$ of the BECs before the loading in the lattice, $S_0/N = 0.72(7)~k_B$ \cite{SuppMat}. This indicates that the transfer from the ODT to the lattice is essentially adiabatic as well.
  

As stated previously, Fig.~\ref{f.thermometry} offers an unbiased calculation of the entropy map of the trapped 3D BH model at fixed particle number. While similar calculations can be found in the literature (for the 1D and 2D BH model \cite{Polletetal2008}, and for the grand-canonical 3D BH model within a mean-field approximation \cite{Yoshimuraetal2008}) such a map for the canonical 3D BH model has not been presented before to the best of our knowledge, and it is therefore worth discussing here. For moderate entropies as those of the experiment ($S/N k_{B} \sim 0.8$), one distinguishes two asymptotic regimes: a SF regime ($u \lesssim 25$) in which the isentropic curves show a slow growth with $u$; and a MI regime (for  $u \gtrsim 35$) in which the isentropic curves grow more rapidly (roughly linearly with $u$). A third intermediate regime separates the SF from the MI regime, in which the isentropic curves show a plateau, compatible with the experimental observations. 
At small $u$, the slow growth of the isentropic curves in the SF regime can be understood within Bogolyubov theory. In a uniform weakly-interacting Bose gas, the Bogolyubov speed of sound $c \propto \sqrt{u}$ increases with $u$, leading to a decrease of the density of states. The temperature dependence of the entropy is $\sim T^3/u^2$, implying that isentropic curves at $S/N k_{B}=s_0$ in the uniform case should grow as $T \sim s_0^{1/3} u^{2/3}$ (within the energy range in which the dispersion relation can be approximated as $\omega(k) = ck$). On the other hand, in the MI regime the entropy of a uniform system with commensurate filling goes as $S/Nk_{B} \sim \exp(-\Delta/T)$ where $\Delta \sim u$ (for $u \gg u_c$) is the MI gap, implying $T \sim u$ along isentropic curves. Note that in the presence of a trap, the cloud wings with $n<1$ evolve towards a hardcore-boson regime, in which the thermodynamics becomes independent of $u$. The intermediate plateau regime looks somewhat unexpected on the basis of these two limiting cases, but it can be understood as a competition between the hardening of the Bogolyubov (phase) mode and the softening of the amplitude mode. The latter indeed becomes gapless at the SF/MI transition and provides a new contribution to the low-energy density of states. A detailed study of the role of the amplitude mode in the thermodynamics will be the subject of future work.


In conclusion, we have estimated the temperature of lattice gases realizing the 3D Bose-Hubbard model from a systematic comparison between the measured momentum-space densities and large-scale unbiased quantum Monte Carlo results. This approach was used across all relevant regimes of the phase diagram. We find temperatures consistent with the preparation of equilibrium states at constant entropy $S/N = 0.8(1) k_{B}$ for all lattice depths. Our results thus indicate that the adiabatic preparation of \emph{finite-entropy} states in quantum simulators is a rather robust property, as the adiabatic nature of the loading process appears to be unaffected by the gapless nature of the excitation spectrum in the superfluid regime, and by the presence of the superfluid/Mott-insulator quantum critical point. This stands in contrast with systematic deviations from adiabaticity (e.g. following the Kibble-Zurek scenario) \cite{Braunetal2015,Keeslingetal2019} which are expected when working at zero entropy. Our findings suggests that ultracold bosons at lower entropies than the ones achieved here (see e.g. Ref.~\cite{Yangetal2020}), combined with flat trapping potentials minimizing finite-size effects, can be adiabatically prepared in the quantum-critical regime of the superfluid/Mott-insulating transition \cite{Sachdev-book}, which still remains largely unexplored using ultra-cold atoms. 

\vspace{0.5cm}
\begin{acknowledgments}
We thank H. Cayla and M. Mancini for their contributions in the early stage of the experiment, as well as A. Browaeys, M. Cheneau and A. Dareau for a critical reading of the manuscript. We acknowledge fruitful discussions with A. Ran\c{c}on and all the members of the Quantum Gas group at Institut d'Optique. All the numerical simulations were performed on the PSMN facilities at the ENS of Lyon. We acknowledge financial support from the LabEx PALM (Grant number ANR-10-LABX-0039), the R\'egion Ile-de-France in the framework of the DIM SIRTEQ, the ``Fondation d'entreprise iXcore pour la Recherche", the Agence Nationale pour la Recherche (Grant number ANR-17-CE30-0020-01). D.C. acknowledges support from the Institut Universitaire de France.
\end{acknowledgments}
  
\bibliography{Bib-Carcy2020}

\begin{widetext}
\begin{center}
\textbf{\large Supplemental Material: Certifying the adiabatic preparation of ultracold lattice bosons in the vicinity of the Mott transition}
\\
\vspace{5mm}
C\'ecile Carcy, Ga\'etan Herc\'e, Antoine Tenart, Tommaso Roscilde, and David Cl\'ement
\vspace{5mm}
\end{center}
\end{widetext}

\setcounter{figure}{0} 
\setcounter{equation}{0} 
\renewcommand\theequation{S\arabic{equation}} 
\renewcommand\thefigure{S\arabic{figure}}  

\section{Microscopic parameters used in the QMC simulations} 

We provide below a list of the microscopic parameters used in the QMC simulations. Here $s$ is the lattice amplitude expressed in units of the recoil energy, and $V_{j}=m \omega_{j}^2 d^2/2$ is the energy offset between adjacent lattice sites induced by the trapping frequency $\omega_{j}/(2\pi)$ at the trap center (and expressed in units of the tunnelling $J$).
\\

\begin{center}
\begin{tabular}{lllll}
 \hline
   s [$E_{R}$] & $U/J$ & $V_{x}$ [$J$] & $V_{y}$  [$J$] & $V_{z}$  [$J$] \\ 
     \hline
    7.75 & 5   &  0.0282  & 0.0252  & 0.0252  \\
    10 & 10.5  &   0.0620 &  0.0556 & 0.0556  \\
    11.05 & 15  &  0.0870 & 0.0780 & 0.0780  \\
    12 & 20  &  0.117  & 0.104  & 0.104 \\
    13 & 25  &   0.157 &  0.140 & 0.140  \\
    13.5 & 30  & 0.181  & 0.162 & 0.162 \\
    14 & 35  &  0.208 &  0.187 & 0.187  \\
    15 & 45  &  0.274  & 0.246  & 0.246  \\
    16 & 60  &  0.357  &  0.320 & 0.320  \\
    17.8 & 92  &  0.564  &  0.506 & 0.506  \\
     \hline
\end{tabular}
\newline
\end{center}

\section{Analysis of the temperature estimate}

As explained in the main text and illustrated in Fig.~\ref{fig3}, the estimate of the temperature is obtained by matching the experimental normalized momentum-space density $\tilde{\rho}_{\rm exp}(k)=\rho(k,0,0)/\rho(0)$ with the theoretical ones $\tilde{\rho}_{\rm QMC}(k;T)= \rho_{\rm QMC}(k,0,0;T)/\rho_{\rm QMC}(0;T)$. To extract the temperature $T$ we build a reduced chi-square quantity as follows,
\begin{equation}
\chi^2_{\rm r}(T)=\frac{1}{N_{p}} \sum_{j=1}^{N_{p}} \frac{[\tilde \rho_{\rm exp}(k_j)- \tilde \rho_{\rm QMC}(k_j;T)]^2}{\sigma_{\rm exp}(k_j)^2}
\label{eq:chiSq}
\end{equation}
where we have discretized the first Brillouin zone with a uniform mesh of $N_{p}=120$ points, and $\sigma_{\rm exp}(k_j)=\sqrt{\tilde \rho_{\rm exp}(k_j)/M}$ is the error estimate on the experimental density (assumed to follow a Poissonian statistics) with $M$ the number of runs of the experiment we use to evaluate $\tilde{\rho}_{\rm exp}(k)$.

\begin{figure}[ht!]
\includegraphics[width=\columnwidth]{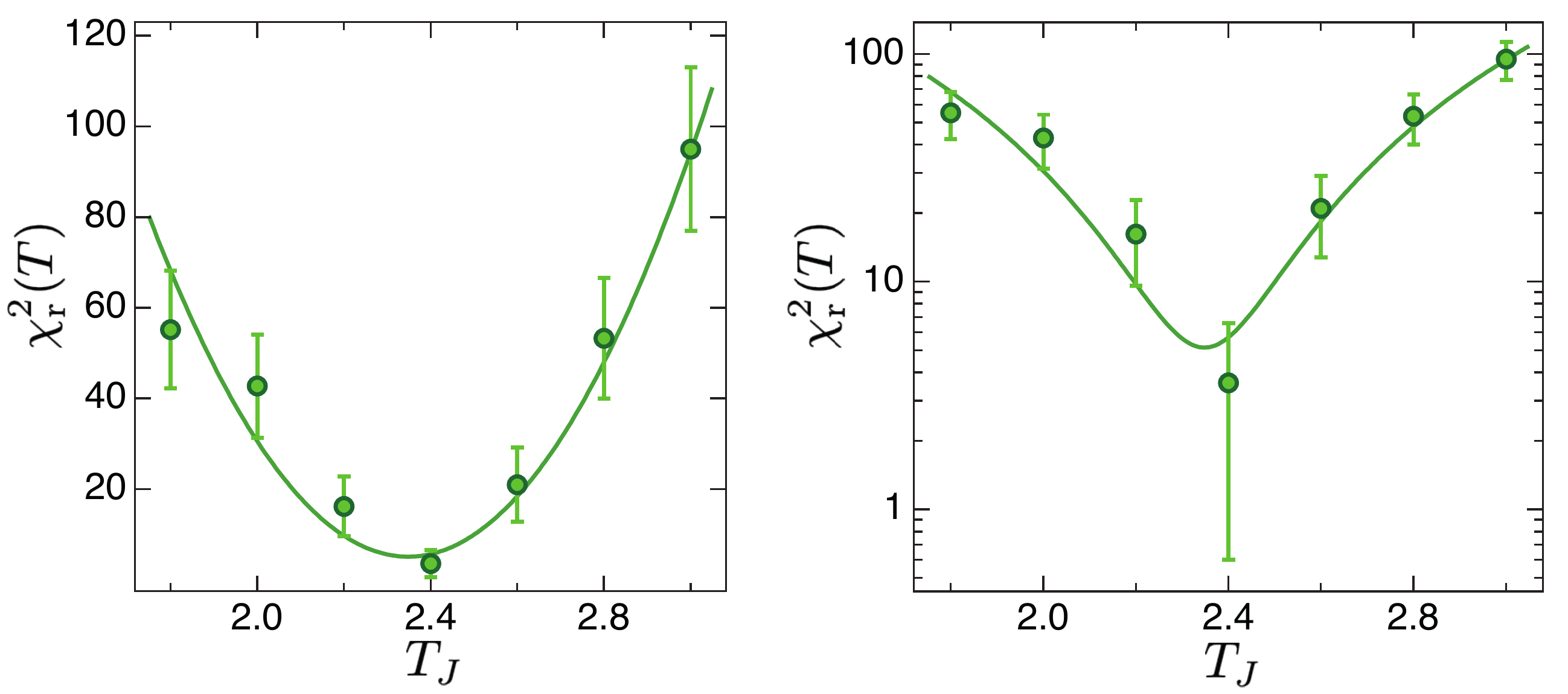}
\caption{The reduced chi-square $\chi_{\rm r}^2$ defined in Eq.~\ref{eq:chiSq} is evaluated for the experimental data set at $u=30$ and for the QMC simulations performed at $u=30$ for various temperatures $T$. A minimum of $\chi_{\rm r}^2$ is clearly identified as a function of $T$, whose position determines the temperature of the lattice gas in the experiment. The solid line is a parabolic fit of the minimum, yielding $T_{J}=2.36(3)$. At $T_{J}=2.4$ we find $\chi_{\rm r}^2=3.6 \pm 3.0$ compatible with unity.}
\label{fig-sup1}
\end{figure}

In Fig.~\ref{fig-sup1}, we show the $\chi^2_{\rm r}$ analysis performed for the experimental data recorded at $u=30$ in the critical regime of the Mott transition. Firstly, we observe a clear minimum of $\chi^2_{\rm r}$ as a function of the temperature, whose position indicates the temperature of the lattice gas in the experiment, {\it i.e.} the temperature which matches best the QMC simulations at the corresponding value $u$. Secondly, we find that the value $\chi_{\rm r}^2(T_{J}=2.4)=3.6 \pm 3.0$ close to the minimum is compatible with unity. This demonstrates that the QMC simulations describe accurately the experimental data within statistical uncertainty. Finally, we fit the dependency of $\chi^2_{\rm r}$ with $T$ with a parabolic profile from which we extract the position of the minimum ($T_{J}=2.36(3)$ for the data set of Fig.~\ref{fig-sup1}). This provides an estimate of the minimal error on the estimated temperature with our approach, which we can compare with the expected limit using the Fisher Information (see below). Note however that the error bars in the main text are larger than those found with this procedure: in order to avoid relying on a fit of $\chi^2_{\rm r}$ to extract the error, the error bars we indicate in Fig.~1 of the main text correspond to the temperature interval over which distinct values of $\chi^2_{\rm r}$ are observed (\emph{e.g.} we use $T_{J}=2.4(2)$ for the data set in Fig.~\ref{fig-sup1}).

\section{Fisher information for temperature estimation based on the momentum-space density}

When using finite statistics, the precision on the estimation of the temperature as a parameter of a distribution (specifically, the $k$-space cut $\tilde \rho_{\rm QMC}(k;T)$ along the $(k,0,0)$ direction) is set by the Fisher information \cite{Fisher-book}
\begin{equation}
I(T) = \sum_k ~\frac{\tilde \rho_{\rm QMC}(k ;T)}{{\cal N}_T} \left [ \frac{\partial \log  \left ( \tilde\rho_{\rm QMC}(k ;T)/{\cal N}_T \right ) }{\partial T_J} \right ]^2 
\end{equation}
where ${\cal N}_T$ is the normalization of  $\tilde \rho_{\rm QMC}(k ;T)$ when summed over $k$. $k$ forms a discrete grid in momentum space -- in our QMC calculations the grid spacing was $\Delta k = k_d/(4L)$ with $L=30$ or $36$, sufficiently fine to capture all the relevant features of the $k$-space density without any sensitive limitation in resolution. 

The Cram\'er-Rao bound on parameter estimation \cite{Fisher-book} stipulates that, when estimating the temperature $T$ using $M$ samples of the momentum distribution, the minimum uncertainty $\delta T$ is set by
\begin{equation}
 \delta T_{J}  = \frac{k_B ~\delta T}{J}    \geq ( \delta T_{J})_{\rm min} = \frac{1}{\sqrt{I(T) M}}~. 
\label{Eq:minT-Fisher}
\end{equation}
In other words, the Fisher information captures the sensitivity of the density $\tilde \rho_{\rm QMC}(k ;T)$ to variations of $T$ and, conversely, the ability that one has to resolve the temperature $T$ by looking at $\tilde\rho_{\rm QMC}(k ;T)$ with finite statistics. 
While the above theoretical minimum may not be attained by comparing the experimental momentum distribution to the theoretical one (see also below the consideration on the particle-number fluctuations), the value of $I(T)$ is an important indication of the potential accuracy of thermometry based on the momentum distribution across the various regimes of the Bose-Hubbard model. 

We have extracted this quantity from the QMC data at fixed particle number $N=3000$: the logarithmic derivative entering in the definition of $I(T)$ has been estimated by the finite difference  
\begin{eqnarray}
\frac{\partial \log  \left ( \tilde\rho_{\rm QMC}(k ;T)/{\cal N}_T \right ) }{\partial (T_J)} \approx \nonumber \\ \frac{1}{\Delta T} \log  \left ( \frac{\tilde\rho_{\rm QMC}(k ;T+\Delta T)}{\tilde\rho_{\rm QMC}(k ;T)} \frac{{\cal N}_{T}}{{\cal N}_{T+\Delta T}} \right )& &
\end{eqnarray}
where $\Delta T = 0.1 J$. The results, presented in Fig.~3 of the main text, show that $I(T)$ has a dramatic excursion of more than 4 orders of magnitude across the phase diagram explored by the experiment, and it reaches the highest values in the temperature range in which the superfluid part of the atomic cloud undergoes its transition to a normal gas. This part is represented primarily by the trap core at $n\approx 1$ when $u \lesssim 30$; beyond this value, the cloud core becomes Mott insulating, and superfluidity is only supported in the trap wings (with $n<1$), but only up to a significantly lower critical temperature. We observe that the peak in $I(T)$ occurs at the maximum temperature for $u\approx 15$, while the peak temperature decreases for lower values of $u$. QMC calculations show that for $u < 15$ the trap frequency in the experiment falls shorter with respect to the value required to maintain a filling $n \approx 1$ at the trap center. As a consequence the critical temperature, which (sufficiently far from the SF/MI transition) is a monotonic function of the filling, is found to be lower at lower $u$.     

Finally, the lower bound on $\delta T$ set by the Fisher information in Eq.~\ref{Eq:minT-Fisher} can be compared to the estimates found in the experiment. Firstly, from the analysis of the experimental data and thermometry, we find that the uncertainty on the reduced temperature $T_{J}=k_{B} T/J$ is of order $\delta T_{J} \sim 0.3$ in the SF regime while it is of the order of $\delta T_{J} \sim 1.5$ deep in the MI regime. The former corresponds to $I(T)\sim1$ and $( \delta T_{J})_{\rm min}\sim 0.4$, while the latter to $I(T)\sim10^{-3}$ and $( \delta T_{J})_{\rm min} \sim 1.3$. Close to the Mott transition, $u \sim u_{c}$, the Fisher information is $I(T)\sim 7$ and $( \delta T_{J})_{\rm min}\sim 0.02$. From the fitting of the minimum of $\chi^2_{\rm r}$ as described previously, we estimate $\delta T_{J} \sim 0.03$, nearly reaching the limit defined by $I(T)$. The fact that  the uncertainty on the thermometry nearly saturates the minimum predicted by the theoretical Fisher information bound is a strong indication that our tomographic measurement of the $k$-space density allows us to essentially extract all the information on the temperature which can be retrieved from the $k$-space density itself.

\section{Impact of particle-number fluctuations on the momentum distribution and on thermometry} 

\begin{figure*}[ht!]
\includegraphics[width=0.8\textwidth]{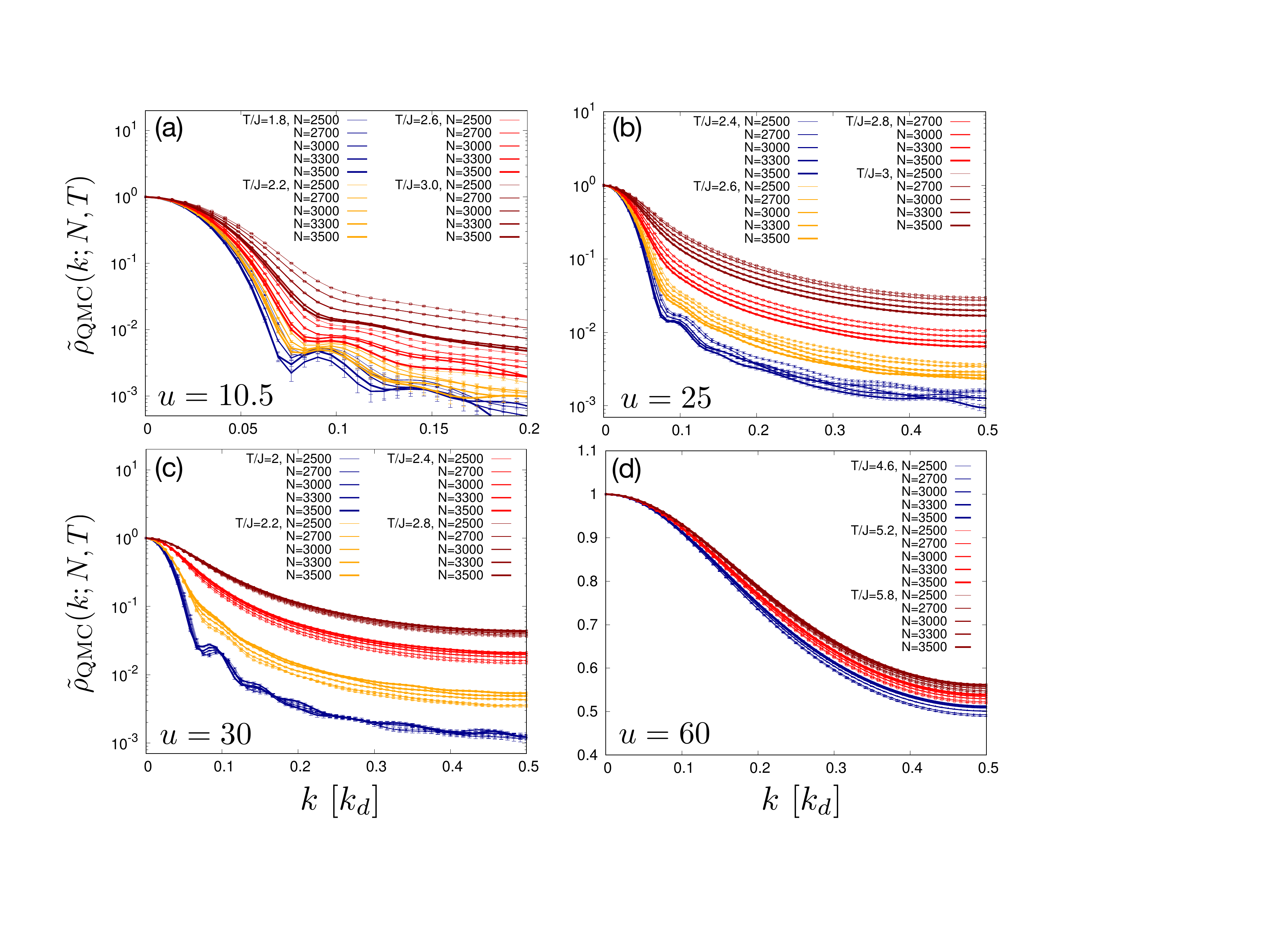}
\caption{{\bf Effect of particle number fluctuations on the normalized momentum-space densities.} The four panels show the peak-normalized momentum distributions $\tilde\rho_{\rm QMC}(k;N,T)$ for different $N$ values in the post-selection interval of the experiment $[2500,3500]$; for various temperatures in the range relevant to the experiment, and for four values of the interaction strength $u$.}
\label{fig-sup2}
\end{figure*}

Quantum Monte Carlo (QMC) simulations allow us to gauge the impact of particle number fluctuations on the $k$-space density  -- and on the associated the $k$-space thermometry -- within the range of $N$ values ($N \in [2500,3500]$) post-selected in the experiment, all the other parameters (temperature, interaction and trapping potential) being fixed. Fig.~\ref{fig-sup1} shows the theoretical quasi-momentum-space density normalized to its peak value, $\tilde{\rho}_{\rm QMC}(k; N,T)$ for various values of $u$ spanning the various regimes of superfluid (SF), normal gas (NG) and Mott insulator (MI). For every temperature, the various $\tilde{\rho}_{\rm QMC}(k ;N,T)$ for different $N$'s span the interval over which one could expect the average experimental data to fall (uniquely because of the uncertainty on $N$) when a temperature $T$ is realized in the experiment. In practice the region spanned by the various $\tilde{\rho}_{\rm QMC}(k ;N,T)$ largely underestimates the range of fluctuations of the averaged experimental data, as it assumes that fluctuations on the averages come uniquely from shot-to-shot variations in $N$, and not from finite statistics (at fixed $N$) nor from other systematic noise sources.   
 
The \emph{a priori} accuracy of thermometry can be fundamentally limited by the variations in $N$. Indeed $\tilde{\rho}_{\rm QMC}(k ;N,T)$ distributions at different temperatures $T$ and $T'$ may overlap when one changes the value of $N$ within the post-selection window of the experiment. In this situation, one could not unambiguously identify the temperature. This aspect offers a further limitation beyond that imposed by the sensitivity of $\tilde{\rho}_{\rm QMC}(k ;N,T)$ to temperature variations, captured by the Fisher information discussed above.  

The temperature grid in the QMC data used to compare with the experimental ones had a step of $(\Delta T)_{\rm min} = 0.2 J$, setting in practice the minimum uncertainty on thermometry. We remind that thermometry has been systematically made by comparing the experimental data with QMC ones at $N=3000$. On the other hand, the experimental fluctuations in $N$ can introduce a higher uncertainty than $(\Delta T)_{\rm min} $, or possible systematic biases, when the curve $\tilde{\rho}_{\rm QMC}(k ;N,T)$ for a given temperature $T$ (used as a reference for thermometry) is actually overlapping with distributions $\tilde{\rho}_{\rm QMC}(k ;N,T')$ at $T' \neq T$. In this case the experimental data with variable $N$ could actually be consistent both with the QMC densities at $T$ and at $T'$. Stated otherwise, a resolution of $\Delta T$ on the estimated experimental temperature $T$ is only meaningful \emph{a priori} when the reference $k$-space density for the thermometry,  $\tilde{\rho}_{\rm QMC}(k ;N=3000,T)$, does not overlap with distributions at $\tilde{\rho}_{\rm QMC}(k ;N',T \pm \Delta T)$ with $N'$ falling within the range of post-selected particle numbers. Our goal below is to assess theoretically the minimal $\Delta T$ satisfying this requirement. 
 
Fig.~\ref{fig-sup2} shows that $N$ fluctuations have substantially different effects on thermometry in the three regimes (SF, NG and MI); and that, in practice, the uncertainty we have quoted on the estimated temperatures is fully consistent with that imposed by particle number fluctuations. We shall review the various regimes and corresponding precision of thermometry below; this discussion offers as well the opportunity to understand the large variations in the Fisher information $I(T)$ across the phase diagram.  
  
\emph{1) SF regime.} Deep in the SF regime ($u = 10.5$ in Fig.~\ref{fig-sup2}(a)) a temperature $T$ significantly smaller than the critical temperature has a small effect on the $k$-space density, as its increase only weakly depletes the condensate -- this justifies the low $I(T)$ in this regime. On the other hand, particle number fluctuations significantly affect $\tilde{\rho}_{\rm QMC}(k ;N,T)$: indeed, when particles are added to the system, a finite fraction thereof goes into the in-trap condensate mode with $k \approx 0$, while the rest spreads rather uniformly over other modes at finite $k$. As a consequence, the momentum tails of $\tilde{\rho}_{\rm QMC}(k ;N,T)$ are significantly suppressed when $N$ increases, due to the renormalization (growing with $N$) by the height of the $k=0$ peak. In the temperature regime of interest to the experiment ($T\approx 2.2 J$) this leads to significant overlap between batches of $\tilde{\rho}_{\rm QMC}(k ;N,T)$ at temperatures differing by $(\Delta T)_{\rm min} = 0.2 J$. A clear separation between the reference curve $\tilde{\rho}_{\rm QMC}(k ;N=3000,T)$ and curves $\tilde{\rho}_{\rm QMC}(k ;N', T \pm \Delta T)$ is only recovered at a lower temperature resolution ($\Delta T = 0.4 J$ -- the one adopted in Fig.~\ref{fig-sup2}(a)), which is consistent with the error bar we estimate for the experimental temperature.  
 
\emph{2) MI regime.} Deep in the MI regime ($u = 60$ in Fig.~\ref{fig-sup2}(d)) temperature effects are uniquely felt by the wings of the cloud as long as $T$ is well below the MI gap for the trap center. In the regime of temperature appropriate for the experiment, $T$ is indeed below the gap but also significantly larger than the critical temperature for the SF/NG transition in the trap wings. Therefore the $k$-space density is very weakly sensitive to temperature changes -- hence the small $I(T)$. Particles added to the system find an incompressible core at the trap center, and therefore they stack primarily on the cloud wings, occupying momentum modes at finite $k$. As a consequence the normalization to the peak height at $k=0$ is nearly independent of $N$, and, unlike what was seen in the SF regime, the wings in $\tilde{\rho}_{\rm QMC}(k ;N,T)$ grow with $N$. This particle number dependence imposes a larger $\Delta T$  ($\Delta T = 0.6 J$ in Fig.~\ref{fig-sup2}(d)) for curves at different temperatures to be discernible; yet in practice the error bar on the experimental temperature is much higher.   
 
\emph{3) Proximity to the SF/NG transition.} In the SF and NG regimes close to the SF/NG transition ($u=25$ and $30$ in Fig.~\ref{fig-sup2}(b-c), thermometry is instead \emph{a priori} optimal, and for two different reasons. On the one hand, the proximity to the SF/NG transition introduces an extreme sensitivity of the momentum distribution to the temperature, resulting in a peak in the Fisher information $I(T)$. This situation alone would already make the effect of particle-number fluctuations much smaller than in the above-cited regimes, given that curves $\tilde{\rho}_{\rm QMC}(k ;N,T)$ and  $\tilde{\rho}_{\rm QMC}(k ;N',T \pm \Delta T)$ are more widely spaced for a given $\Delta T$ than in the regimes discussed above. But a further benign effect adds up, namely the relative insensitivity of the $\tilde{\rho}_{\rm QMC}(k ;N,T)$ densities to $N$ variations in the temperature regime of the experiment. As seen in the two previous examples, the wings of the $\tilde{\rho}_{\rm QMC}(k ;N,T)$ decrease with $N$ in the SF regime, while they increase in the MI regime. Clearly in the transition regime ($u \approx 30$) the $N$ dependence becomes minimal, as one flips from one behavior to the other. Under these circumstances, curve batches separated by $\Delta T = 0.2 J$ are perfectly discernible; therefore an error $0.2 J$ on the estimated temperature is \emph{a priori} meaningful, and in practice it is the one we can attach to the experimental temperature for $u = 30$ (which marks the optimal thermometry point).

In conclusion, the effect of particle-number fluctuations is potentially very important in limiting the accuracy of thermometry beyond what predicted by the Fisher information $I(T)$ at fixed $N$. Yet our estimated error bars on the experimental temperatures are systematically larger than the minimal $\Delta T$ allowed by particle-number fluctuations within the experimental range of $N$ and $T$. Optimal thermometry with minimal impact of particle number fluctuations is achieved in proximity to the SF/NG transition for strong interactions ($u \approx 30$).

\begin{figure}[ht!]
\includegraphics[width=\columnwidth]{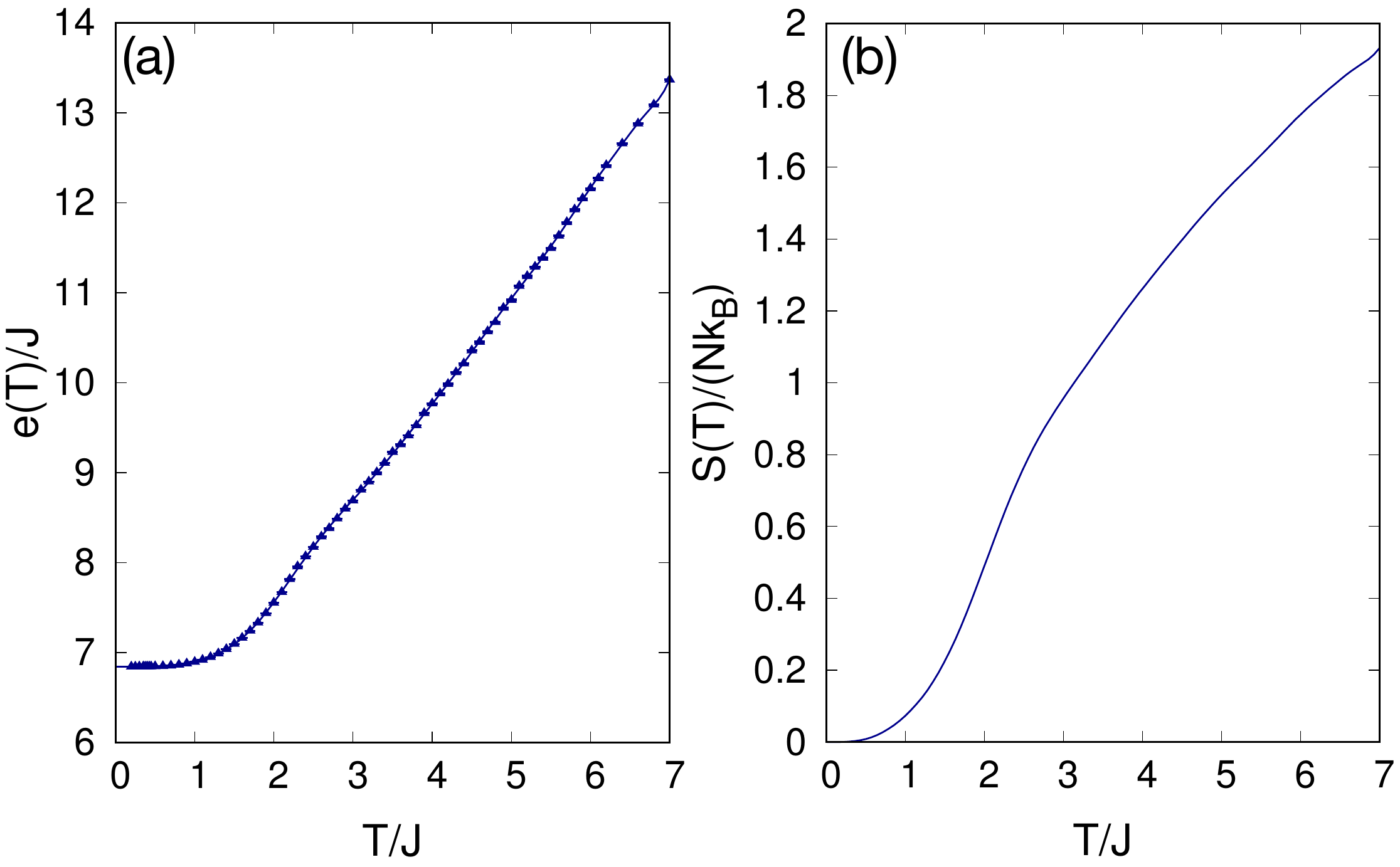}
\caption{(a) Energy per particle for $u=30$ and $N=3000$, calculated with QMC on a $30^3$ box with periodic boundary conditions. The solid line shows a polynomial fit of order $n=22$; (b) Entropy per particle extracted from the fit to the energy.}
\label{fig3-sup}
\end{figure}

\section{Extracting the theoretical entropy curves from the QMC data}

As described in the main text, the entropy is extracted by fitting the QMC data for the energy per particle $e(T) = \langle {\cal H} \rangle/N$ with a high-order polynomial (up to order $n=22$), and obtaining an ``analytical" energy curve $e_{\rm fit}(T)$ which provides a smooth interpolation of the numerical data. The entropy per particle is then extracted as
 \begin{equation}
S(T)/N = \int_0^T \frac{d\theta}{\theta} ~\frac{de_{\rm fit}(\theta)}{d\theta}
 \end{equation}
 admitting a simple analytical expression when $e_{\rm fit}(T)$ is a polynomial in $T$. Fig.~\ref{fig3-sup} shows an example of this analysis for the case $u=30$.

\section{Measured entropy of the 3D BECs before loading in the lattice}

\begin{figure}[ht!]
\includegraphics[width=\columnwidth]{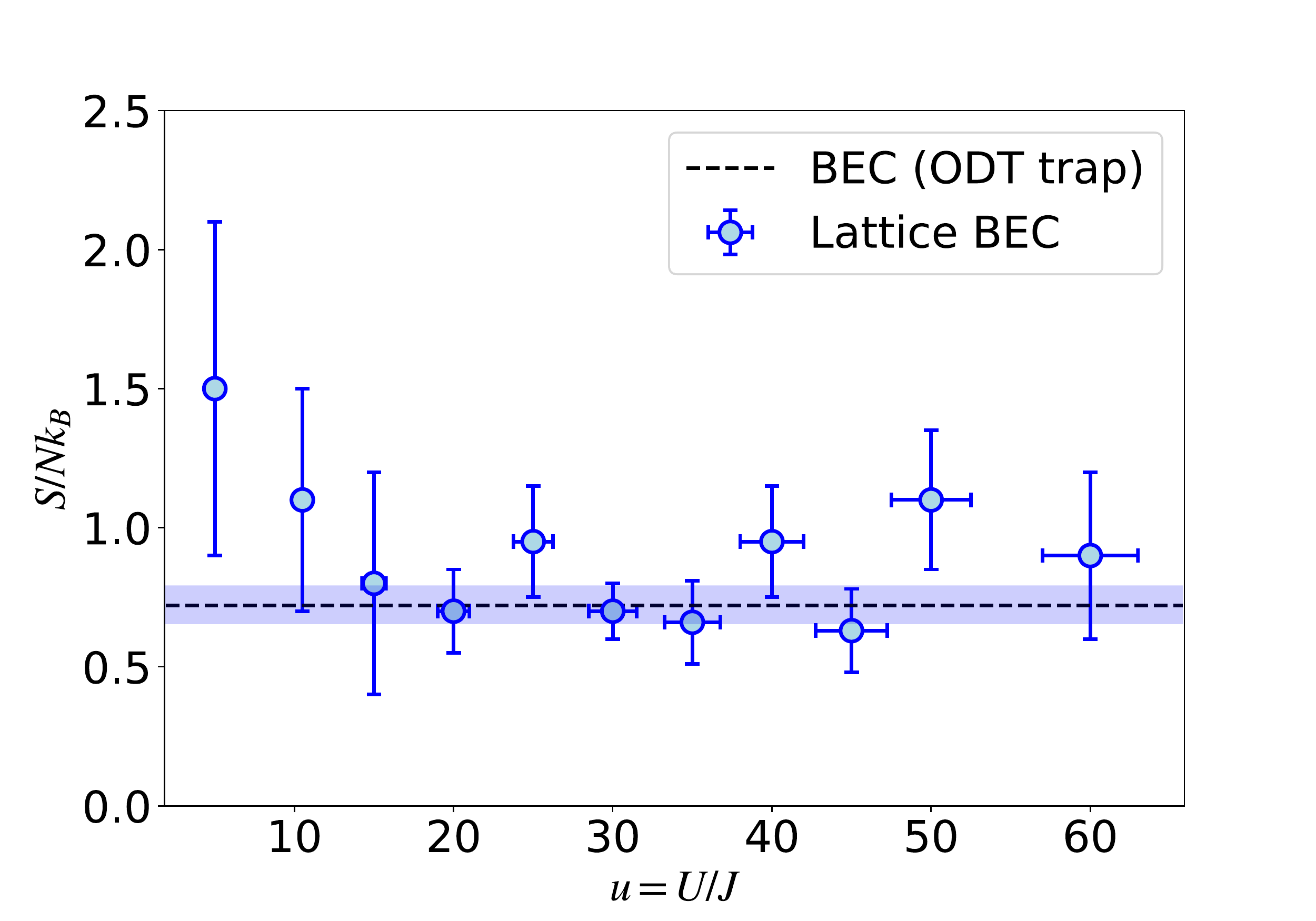}
\caption{Entropy per particle $S/N k_{B}$ of the lattice gas (blue dots) evaluated by comparison with QMC calculations and plotted as a function of $u=U/J$. The dashed line is the measured value of the entropy per particle $S_{0}/N k_{B}$ of the 3D BECs before the loading in the lattice (the shaded area around the dashed line corresponds to one standard deviation).}
\label{fig4-sup}
\end{figure}

To evaluate the entropy $S_{0}$ of the BECs in the optical-dipole trap, before the loading into the optical lattice, we have measured the condensate fraction $f_{c}$ of the gas, and made use of the relation for a non-interacting partially-condensed Bose gas in a harmonic trap \cite{Pitaevskii-book}

\begin{equation}
S_{0}/N k_{B}=\frac{4 g_{4}(1)}{\eta(3)} \ (1-f_{c}),
\end{equation}

where $g_{4}(1)\simeq1.082$ and $\eta(3)\simeq1.2026$.

The condensed fraction $f_{c}$ was extracted from the measured $k$-space density with a by-modal fit of the condensed and non-condensed components. The use of the above relation, valid for a non-interacting Bose gas, is justified by the fact that the thermal energy $k_{B}T$ exceeds the mean-field interaction energy $\mu$. In our experiment, we find $k_{B} T \sim h \times 2380~$Hz $ \gg \mu \sim h \times 350~$Hz. Corrections to the above estimate of $S_{0}$ coming from interactions are therefore expected to be small.

Fig.~\ref{fig4-sup} compares the estimated entropy for the BEC before optical lattice ramp with the entropies estimated at finite optical lattice depth. We observe that in the parameter regime in which thermometry is most accurate, $u\sim u_{c} =29.3$, the agreement between the measured entropies $S_{0}$ of the BEC in the ODT and $S$ of the lattice gas is excellent. 

\end{document}